\documentclass{jaa}

\usepackage{graphicx}

\begin{document}

\title{Relativistic disc line: a tool to constrain neutron star equation of state models}


\author{Sudip Bhattacharyya\textsuperscript{1}}
\affilOne{\textsuperscript{1}Department of Astronomy and Astrophysics, Tata Institute of Fundamental Research, 1 Homi Bhabha Road, Mumbai 400005, India.\\}


\twocolumn[{

\maketitle

\corres{sudip@tifr.res.in}


\begin{abstract}
Relativistic iron K$\alpha$ spectral emission line from the inner disc of a neutron star low-mass 
X-ray binary (LMXB) was first detected in 2007. This discovery opened up new ways to probe strong gravity
and dense matter. The past decade has seen detections of such a line from many neutron star LMXBs,
and confirmation of this line from the same source with several X-ray satellites.
These have firmly established the new field of relativistic disc line from neutron star systems 
in only ten years.  Fitting the shape of such a line with an appropriate general relativistic model
provides the accretion disc inner edge radius to the stellar mass ratio. In this review, we briefly discuss
how an accurate measurement of this ratio with a future larger area X-ray instrument can be used to constrain 
neutron star equation of state models.
\end{abstract}

\keywords{accretion, accretion discs---equation of state---methods: numerical---stars:
neutron---X-rays: binaries.}

}]



\section{Introduction}\label{Introduction}

Knowledge of the cold degenerate matter (e.g., $\sim 10^8$~K) at several
times nuclear saturation density is a fundamental goal of physics. 
Such high-density matter may be in exotic states such as pion or
kaon condensates, or strange quark matter. Such
matter cannot be studied with observations of the early universe or
by heavy-nuclei collision experiments (\"Ozel and Psaltis 2009; van Kerkwijk 2004; 
Blaschke {\em et al.} 2008; Lattimer and Prakash 2007). However, this
super-dense degenerate matter exists in the cores of neutron stars, and plausibly 
the only way to probe it is by ruling out as many theoretically proposed equation of 
state (EoS) models of the stellar core as possible.

But how to constrain EoS models of the deep interior of a neutron star?
While the internal constitution of a neutron star depends on the EoS model,
which connects the total mass-energy density ($\epsilon$) with the pressure ($p$) 
in a degenerate condition, such a model also determines the stellar bulk properties.
For example, the stellar global structure can be computed from the Tolman-Oppenheimer-Volkoff
(TOV) equation:
\begin{eqnarray}
	\frac{{\rm d}p(r)}{{\rm d}r} = -\frac{G}{c^2}\frac{[p(r)+\epsilon(r)][m(r)+4\pi r^3p(r)/c^2]}{r(r-2Gm(r)/c^2)},
	\label{TOV}
\end{eqnarray}
if the star is non-spinning. Here, $m(r) = 4\pi \int_0^r {\rm d}r^{\prime}r^{\prime 2}\epsilon(r^{\prime})$
is the gravitational mass inside a sphere of radius $r$.
Therefore, if one can accurately measure two independent global parameters, such as gravitational mass
(hereafter, mass $M$) and radius ($R$) of a non-spinning star, most of the theoretically proposed EoS models
can be effectively ruled out. For a fast-spinning star, one needs to measure an additional
parameter, e.g., the stellar spin frequency ($\nu$). However, three required parameters have so far not 
been precisely measured for any neutron star. 

Note that the recent precise measurement of the mass ($2.01\pm0.04 M_\odot$) of the
millisecond pulsar PSR J0348+0432 rules out the EoS models which cannot support
such a high mass value (Antoniadis {\em et al.} 2013). But, while this has somewhat constrained 
the parameter space, all types of EoS models, such as nucleonic, strange quark matter, hybrid 
and hyperonic, still survive (Bhattacharyya {\em et al.} 2016; Bhattacharyya {\em et al.} 2017).

\begin{figure*}
	\centering\includegraphics[height=.35\textheight]{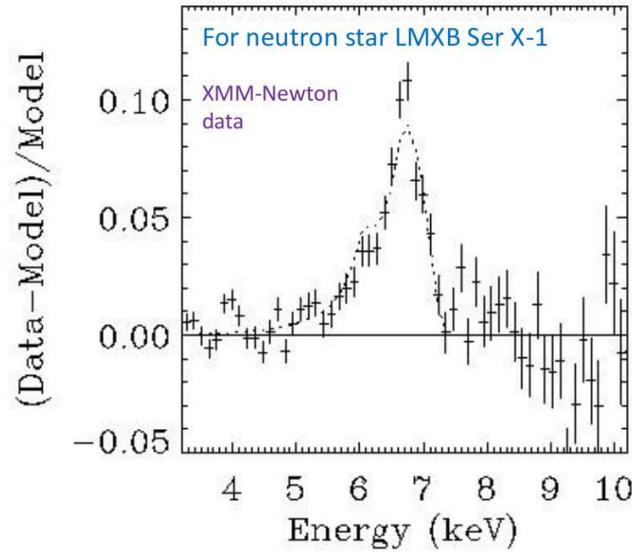}
	\caption{Relativistic iron K$\alpha$ spectral emission line from
	the inner accretion disc of the neutron star LMXB Ser X--1
	({\em XMM-Newton} satellite data; Bhattacharyya and Strohmayer 2007). This line is shown as the
	X-ray intensity in excess to the best-fit
	continuum spectral model and normalized by this continuum model. A broad asymmetric line is clearly
	shown by the data points, which is well-fit with the dotted profile
	of a relativistic model (see Section~1).
\label{f1}}
\end{figure*}

Accurate measurements of three independent global parameters of the same neutron star is extremely
difficult because of a number of unknown systematics. However, such systematic uncertainties can be reduced
if complementary methods to measure the stellar parameters are available for a given neutron star.
A low-mass X-ray binary (LMXB), in which a neutron star accretes matter from a Roche-lobe filling
low-mass companion (Bhattacharyya 2009), can provide several such complementary methods, and hence
can be a particularly promising system for parameter measurements (see Bhattacharyya 2010 for a review).
These methods involve various spectral and timing features, such as thermonuclear X-ray bursts and their
various properties, accretion-powered pulsations, kilo-Hertz quasi-periodic oscillations (kHz QPOs), broad
relativistic iron lines, quiescent emissions and timing features due to orbital motion, as 
complementary tools to reliably measure neutron star parameters (Bhattacharyya 2010). Here, we focus on one of these
methods, the one involving the broad relativistic iron line.

A broad relativistic iron K$\alpha$ spectral emission line is observed
from many stellar-mass and supermassive black hole systems
(Fabian {\em et al.} 2000; Reynolds and Nowak 2003; Miller 2007).
Such a fluorescent line near 6 keV is believed to originate from the
reflection of hard X-rays from the geometrically thin accretion disc.
The hard X-ray source could be an accretion disc corona (Fabian {\em et al.} 2000),
or even the base of a jet (Markoff and Nowak 2004).
An incident X-ray photon can either be Compton scattered by
free or bound electrons, or be subject to photoelectric absorption.
Such an absorption is followed by either Auger de-excitation or fluorescent 
line emission (Fabian {\em et al.} 2000). While many such lines are generated,
the strongest among them is the one for the $n = 2 \rightarrow n = 1$ transition of the iron atom or ion.
Such an intrinsically narrow iron line in the range $6.4-6.97$~keV is broadened and shaped by
various physical effects, e.g., Doppler effect, special
relativistic beaming, gravitational redshift and general
relativistic light-bending (Tanaka {\em et al.} 1995). 
Fitting the shape and energies of this line with an appropriate 
relativistic model can, therefore, be used to measure the inner-edge radius
$r_{\rm in}$ of the accretion disc in the unit of the black hole mass $M$. By
assuming that the disc extends up to the innermost stable circular orbit (ISCO),
i.e., $r_{\rm in}$ is the ISCO radius $r_{\rm ISCO}$,
one can infer the black hole angular momentum parameter $a$ from the known 
expression of $r_{\rm ISCO}c^2/GM$ for the Kerr spacetime (Miller 2007).
Here, $a = Jc/GM^2$, with $J$ be the black hole angular momentum.

Note that $r_{\rm in}$ can evolve with accretion rate for a
source, as the disc can be truncated at various radii due to magnetic field,
radiative pressure or other reasons. 
Therefore, the ISCO radius $r_{\rm ISCO}$ is the lower limit of $r_{\rm in}$
for an accreting black hole. Hence, the minimum inferred value of $r_{\rm in}c^2/GM$
for a given black hole source can be considered an upper limit of $r_{\rm ISCO}c^2/GM$, which
gives the lower limit of the angular momentum parameter $a$ of that source.

If the relativistic disc line were also observed from an accreting neutron star, then that 
could be used as a tool to measure stellar parameters, and hence to constrain EoS models.
But, while  a broad iron line from neutron star LMXBs had been known (see, for example, Asai et al. 2000),
their characteristic asymmetry, and hence the relativistic nature and the inner disc origin,
could not be established till 2006. In 2007, Bhattacharyya and Strohmayer (2007) established
the inner accretion disc origin of the broad iron line from a neutron star LMXB for the first time
(see Figure~\ref{f1}). This was done by analyzing the {\it XMM-Newton} satellite data from Serpens X--1 (Ser X--1),
which opened a new area of science and supported the continuation of this satellite 
(McBreen and Schartel 2008).

In Section~2, we outline the rapid development of the field of relativistic lines from
neutron star LMXBs during the past decade. In Section~3, we briefly describe how such a spectral line can be used to
constrain stellar EoS models. Finally, we briefly mention our conclusions in Section~4.

\section{Relativistic iron spectral line from neutron star LMXBs}\label{Details}

Soon after the first detection of broad relativistic iron spectral emission line from the
{\em XMM-Newton} data of Ser X--1 (Bhattacharyya and Strohmayer 2007), this line was
confirmed from the same source with another satellite {\em Suzaku} (Cackett {\em et al.}, 2008).
This paper also reported such lines from the {\em Suzaku} data of two other neutron star
LMXBs: 4U 1820--30 and GX 349+2, and found that the measured inner disc radii from iron lines
of these two sources are consistent with the inner disc radii implied by the high-frequency 
timing feature kHz QPOs. Next, Pandel {\em et al.} (2008) reported the detection of a broad 
relativistic iron line from the {\em XMM-Newton} and {\em RXTE} satellite data of 4U 1636--536. 
This line feature indicated a combination of several iron K$\alpha$ lines in different ionization states.
Papitto {\em et al.} (2009) reported a highly significant iron K$\alpha$ emission line from the
{\em XMM-Newton} data of the accretion-powered millisecond pulsar SAX J1808.4--3658. Assuming a stellar mass of $1.4~M_\odot$,
the inferred value of $r_{\rm in}$ was $18.0_{-5.6}^{+7.6}$~km. This was less than the pulsar
corotation radius ($31$ km), which was consistent with the fact that the accretion-powered pulsations
were observed during the iron line detection. Di Salvo {\em et al.} (2009) found not only 
a broad relativistic iron emission line from the {\em XMM-Newton} data of the neutron star LMXB
4U 1705--44, but also other broad low-energy emission lines from this source. These
low-energy lines and the iron line appeared to be produced from the same inner disc region.

\begin{figure*}
	\centering\includegraphics[height=.65\textheight,angle=-90]{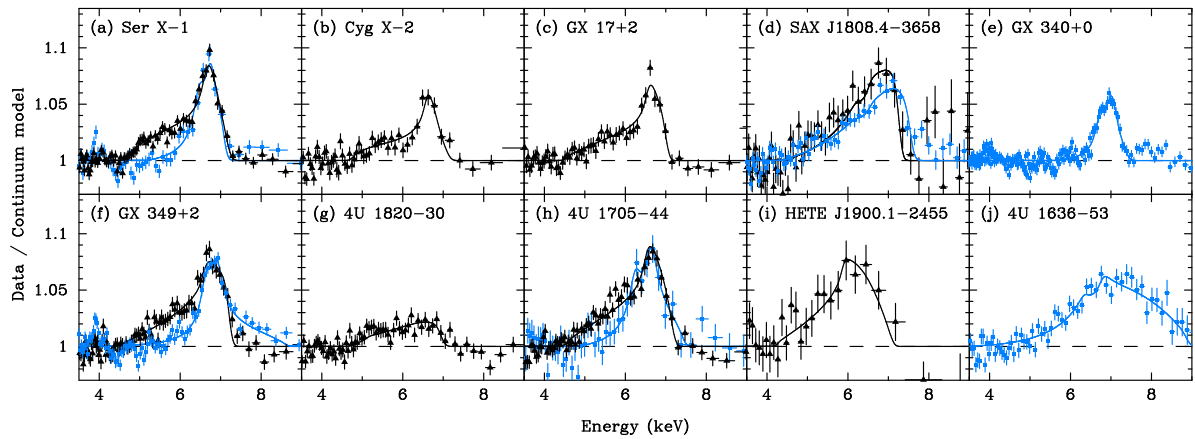}
	\caption{Examples of broad relativistic iron K$\alpha$ spectral emission line from 
	a number of neutron star LMXBs. Here, data to the continuum model ratio is shown. Best-fit
relativistic models are shown by solid lines. {\em Suzaku} satellite data are displayed in black and
the {\em XMM-Newton} satellite data are shown in blue (Section~2; figure courtesy:
E. M. Cackett; Cackett {\em et al.} 2010).
\label{f2}} 
\end{figure*}

\begin{figure*}
	\centering\includegraphics[height=.35\textheight]{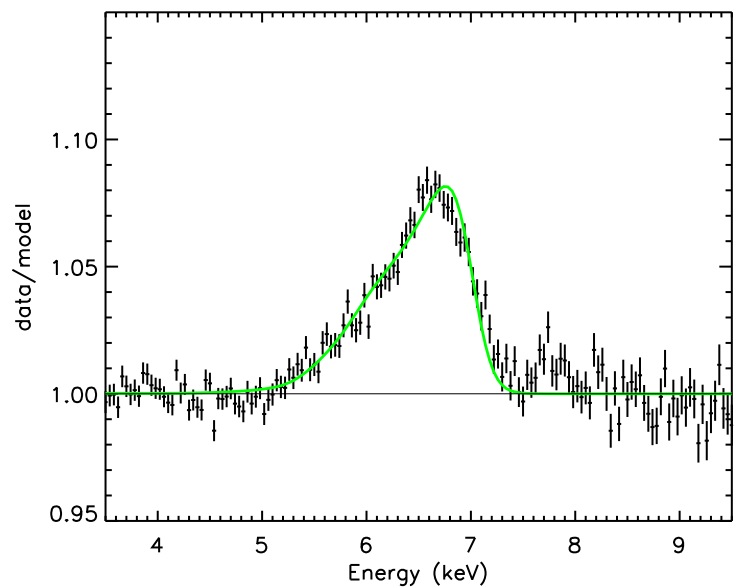}
	\caption{Broad relativistic iron K$\alpha$ spectral line from the {\em NuSTAR}
	data of the neutron star LMXB Ser X--1. Data to continuum model ratio is shown.
The green line shows the best-fit relativistic model (Section~2; figure courtesy:
J. M. Miller; Miller {\em et al.} 2013).
\label{f3}}
\end{figure*}

\begin{figure*}
	\centering\includegraphics[height=.35\textheight]{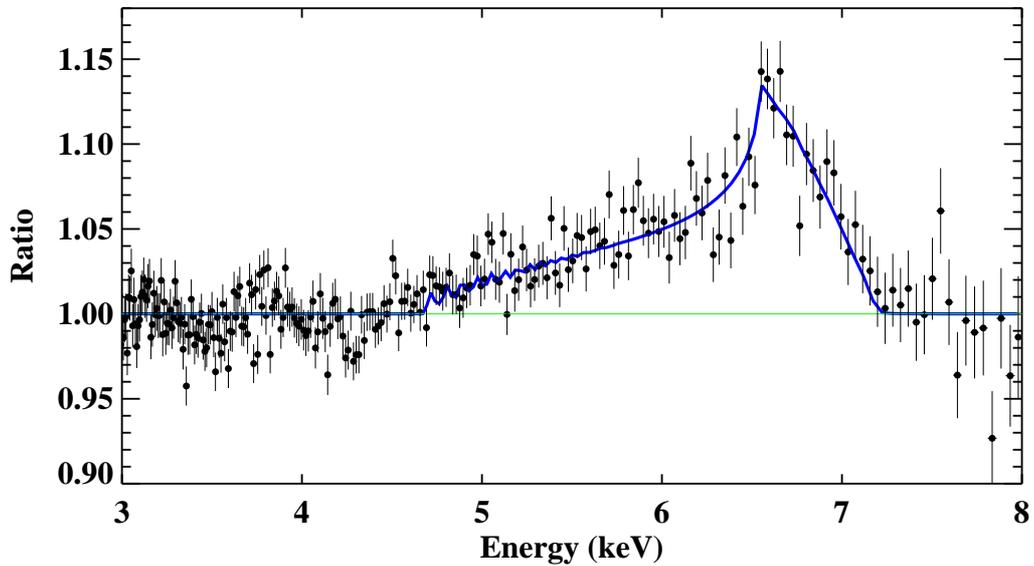}
	\caption{Broad relativistic iron K$\alpha$ spectral line from the {\em Chandra}
	        data of the neutron star LMXB Ser X--1. Data to continuum model ratio is shown.
	The blue line shows the best-fit relativistic model (Section~2; figure courtesy:
	C.-Y. Chiang; Chiang {\em et al.} 2016a).
\label{f4}}
\end{figure*}

\begin{figure*}
	\centering\includegraphics[height=.35\textheight]{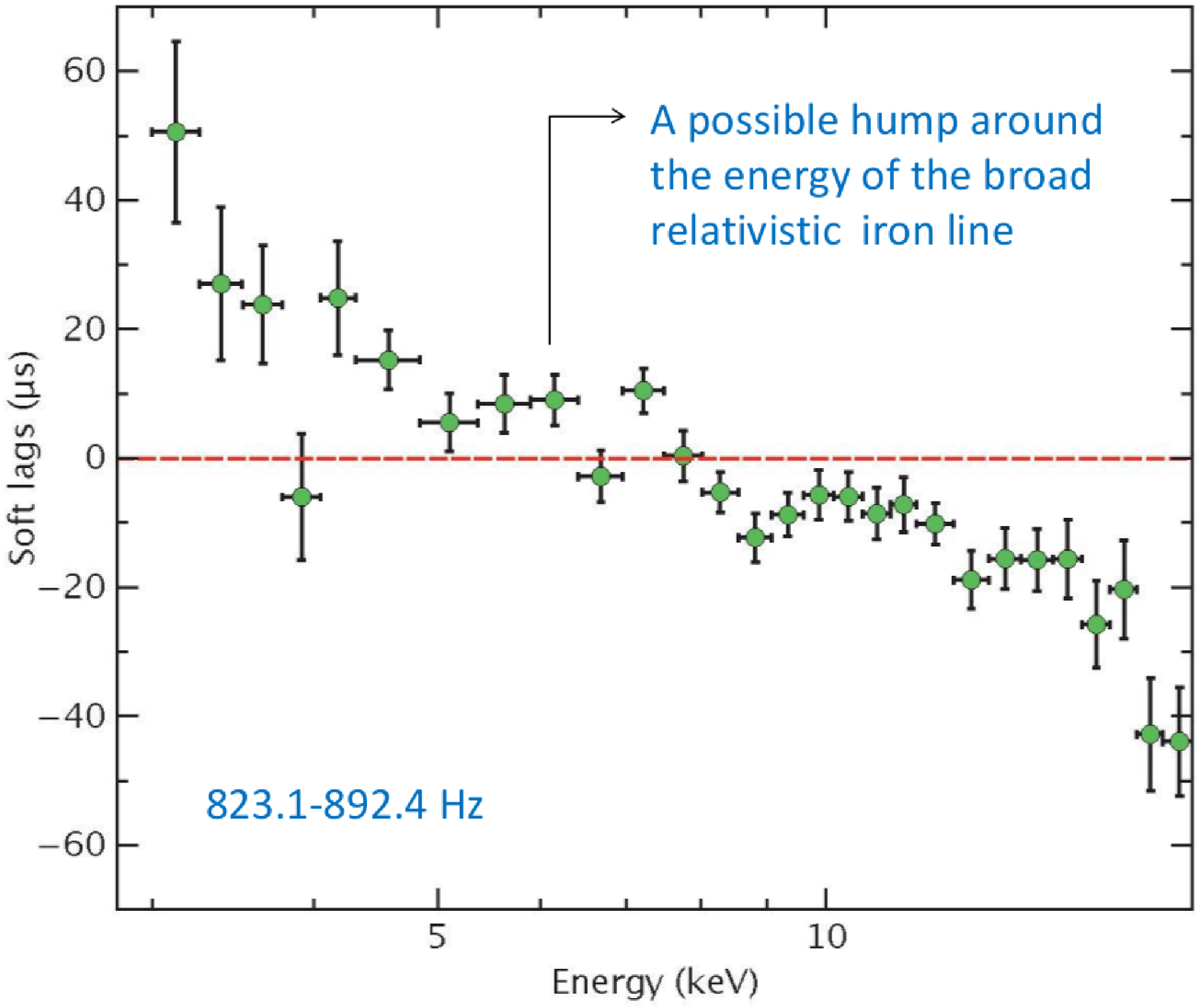}
	\caption{Mean time-lag energy spectrum for the kHz QPO frequency range ($823.1-892.4$~Hz) from the
	{\em RXTE} satellite data of the neutron star LMXB 4U 1608--522 (Barret 2013). A tentative hump
	around $5-7$~keV is suggestive of the reflection origin of the iron line 
	(Section~2; figure courtesy: D. Barret).
\label{f5}}
\end{figure*}

The new field of broad relativistic iron spectral emission line from neutron star systems evolved rapidly, and such a line 
was detected from 10 sources within only three years of the first detection (Cackett {\em et al.} 2010 and
references therein). This paper presented a comprehensive, systematic analysis of {\em Suzaku}
and {\em XMM-Newton} spectra of these sources, of which four are very bright and persistent Z sources,
four are medium bright atoll sources, and two are accretion-powered millisecond pulsars (see Figure~\ref{f2}). 
Cackett {\em et al.} (2010) suggested that the boundary layer of the neutron star illuminates the disc. 
Note that, a boundary layer does not exist in case of an accreting black hole, and a corona or 
the base of a jet provides the illuminating hard X-rays
for black hole sources. Cackett {\em et al.} (2010) reported that $r_{\rm in}c^2/GM = 6-15$ for most cases. Moreover,
although $r_{\rm in}c^2/GM < 6$ values were not explored because Schwarzschild spacetime was used,
the best-fit $r_{\rm in}c^2/GM$ value pegged at the lower limit $6$ for many spectra. This implies that
$r_{\rm in}c^2/GM$ value could be somewhat smaller than $6$ for many cases. However, unlike 
black hole systems, $r_{\rm in}c^2/GM$ cannot be much smaller than $6$, because the disc has to terminate
at the hard surface of the neutron star.

Since the shapes of broad relativistic iron lines were measured with X-ray charge-coupled devices (CCDs),
pile-up could affect such measurements. In case of pile-up, two or more photons are registered as a single event
during a given CCD frame, and thus pile-up can distort the spectrum. Therefore, for a bright source like
a neutron star LMXB, the inferred relativistic nature of the iron emission line could be a result of pile-up.
Hence, Miller {\em et al.} (2010) assessed
the impact of this effect on relativistic disc lines, and found that a severe pile-up falsely narrows the emission lines.
Therefore, the observed large width of such a line cannot be a result of the pile-up.
Moreover, when the pile-up effect is modest, the relativistic disc spectroscopy is not significantly affected
(see, for example, Figure~11 of Miller {\em et al.} 2010).

When the {\em NuSTAR} satellite was launched, Ser X--1 was again observed with this satellite for two reasons:
(1) the broadband capability (in $3-79$~keV) of {\em NuSTAR} could characterize the source continuum spectrum
in an unprecedented manner, and (2) the ability of {\em NuSTAR} to measure the iron K$\alpha$ line free of 
photon pile-up distortions (Miller {\em et al.} 2013). The relativistic nature of the broad iron K$\alpha$ line
detected with {\em NuSTAR}, 
which suggested a disc inner edge close to the ISCO, was $5\sigma$ significant (see Figure~\ref{f3}). Moreover, a broad hump around 
$10-20$~keV in the continuum spectrum was detected for the first time from a neutron star LMXB.
Such hump is expected from Compton back-scattering, and was earlier detected from a number of black hole systems.
As if the confirmation of the relativistic disc line from three satellites were not enough, Ser X--1 was observed
once again with the {\em Chandra} satellite (Chiang {\em et al.} 2016a). 
This longest (300 ks) observation with the ``High Energy Transmission 
Grating Spectrometer" in the ``continuous clocking" mode was free of the pile-up effects and would detect 
narrow lines in the iron K range, if present. As expected, the relativistic nature of the line 
was clearly found (see Figure~\ref{f4}). However, there was no strong evidence of narrow lines.

If the broad iron line is produced by the reflection of hard X-ray photons from the disc, then
this line is expected to be delayed with respect to the illuminating hard X-ray source by the 
light travel time between this illuminating source and the disc. This delay could be manifested
in the time-lag spectrum of kHz QPOs, which are believed to be produced close to the accreting 
neutron star. Indeed, Barret (2013) reported a clear lag of $3-8$~keV photons with respect to $8-30$~keV
photons for a range of kHz QPO frequencies. 
The amount of this lag ($\sim 15-40$~$\mu$s) is consistent with the light travel time for a distance 
of a few Schwarzschild radii, that is a rough distance between the neutron star and the inner disc.
Moreover, a tentative hump in the time-lag spectrum around the broad iron K$\alpha$ line 
energy indicates that this line is a part of the delayed reflected spectrum (see Figure~\ref{f5}).
This, the observations of relativistic disc line from the same source with several satellites, and 
further detections and measurements of such lines from neutron star LMXBs with {\em NuSTAR} 
and other satellites (e.g., Chiang {\em et al.} 2016b; Ludlam {\em et al.} 2017a; 
Ludlam {\em et al.} 2017b; Mondal {\em et al.} 2017) have firmly established this new field within
a decade from the first detection.

\section{Ways to constrain EoS models}\label{Ways}

In this section, we discuss how the dimensionless inner edge radius $r_{\rm in}c^2/GM$ of the 
geometrically thin, Keplerian disc, inferred from fitting the shape of a relativistic disc line, 
can be used to constrain neutron star EoS models. As mentioned in Section~1,
measurements of three independent stellar parameters are required to probe the EoS models.
Here, we propose to use $r_{\rm in}c^2/GM$ as one of these three parameters. More specifically,
our method involves constraining the $M$ versus $r_{\rm in}c^2/GM$ space and/or the $Rc^2/GM$ versus 
$r_{\rm in}c^2/GM$ space in order to constrain EoS models. This is because $M$ and $Rc^2/GM$
could be independently inferred for some neutron star LMXBs. 
For example, $M$ could be measured by the binary
orbital motion method (e.g., Steeghs and Casares 2002; Bhattacharyya 2010), while $Rc^2/GM$ could be inferred using thermonuclear 
burst oscillations (e.g., Bhattacharyya {\em et al.} 2005). However, in order to rule out EoS models
from the observational constraints on the above mentioned parameter spaces, one needs
to theoretically compute $r_{\rm in}c^2/GM$ for various EoS models.

\begin{figure*}
	\centering\includegraphics[height=.25\textheight]{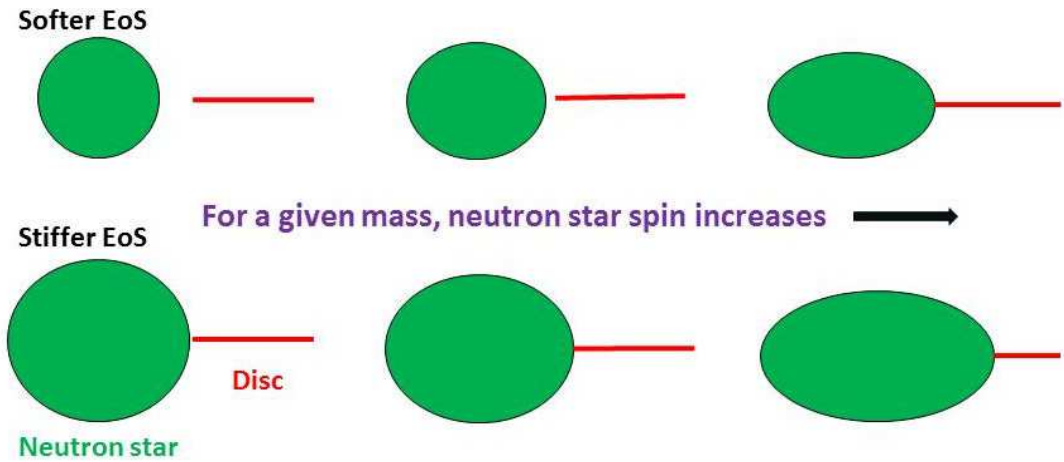}
	\caption{A cartoon showing the effects of neutron star EoS and spin rate on the location of 
		the accretion disc inner edge. Here we assume that the disc terminates at the stellar surface
		or at the ISCO, whichever is bigger. For the same mass and spin rate, the neutron star is 
		smaller for a softer EoS. In case of such an EoS and a smaller spin rate, it is more likely that 
		the disc would not touch the stellar surface and would terminate at the ISCO. For a higher
		spin rate and the same mass and EoS, the stellar equatorial radius would increase and the
		(corotating) disc inner edge radius would likely decrease. But when the stellar 
		equatorial radius becomes large enough for a even higher spin rate, the disc would terminate 
		at the stellar surface instead of the ISCO radius, and the disc inner edge radius would naturally
		increase for a further increase of the spin rate. For a stiffer EoS model, the qualitative 
		behaviour of the disc inner edge radius is the same, but, as the neutron star is bigger 
		for this case, the disc could touch the star for a smaller spin rate. Note that the stellar
		deformation due to spin is exaggerated in this figure (see Section~3).
	\label{f6}}
\end{figure*}

\begin{figure*}
	\centering\includegraphics[height=.35\textheight]{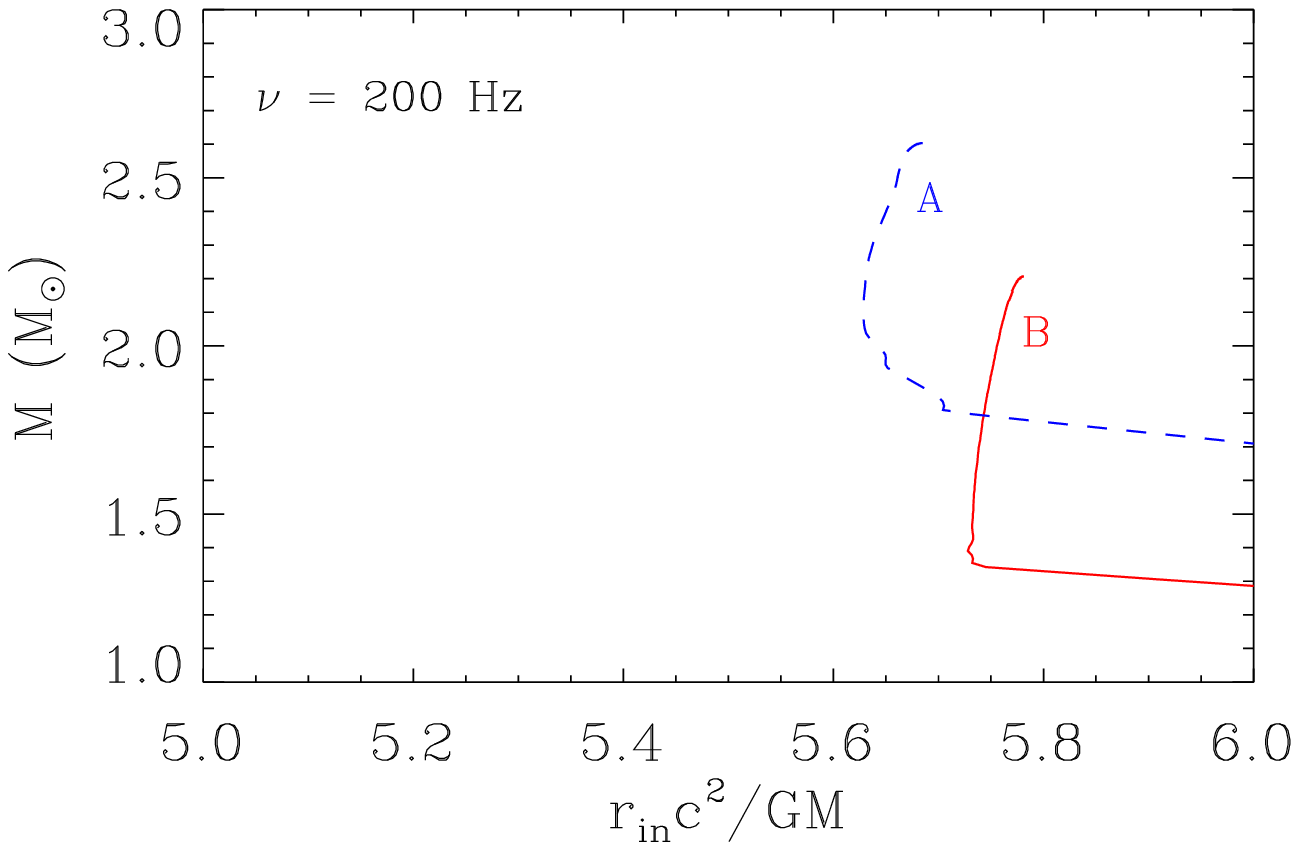}
	\caption{Neutron star mass versus disc inner edge radius-to-stellar-mass ratio for
	two realistic EoS models ({\em A} and {\em B}; see Section~3) for a 
	stellar spin frequency of 200 Hz (Bhattacharyya 2011).
	The straight line portions of the curves correspond to the disc termination on the stellar surface.
	This figure shows how a measurement of the disc inner edge radius-to-stellar-mass ratio
	from the fitting of a broad relativistic iron K$\alpha$ spectral line can be used to constrain
	neutron star EoS models, when the stellar spin frequency is known.
	\label{f7}}
\end{figure*}

\begin{figure*}
	\centering\includegraphics[height=.35\textheight]{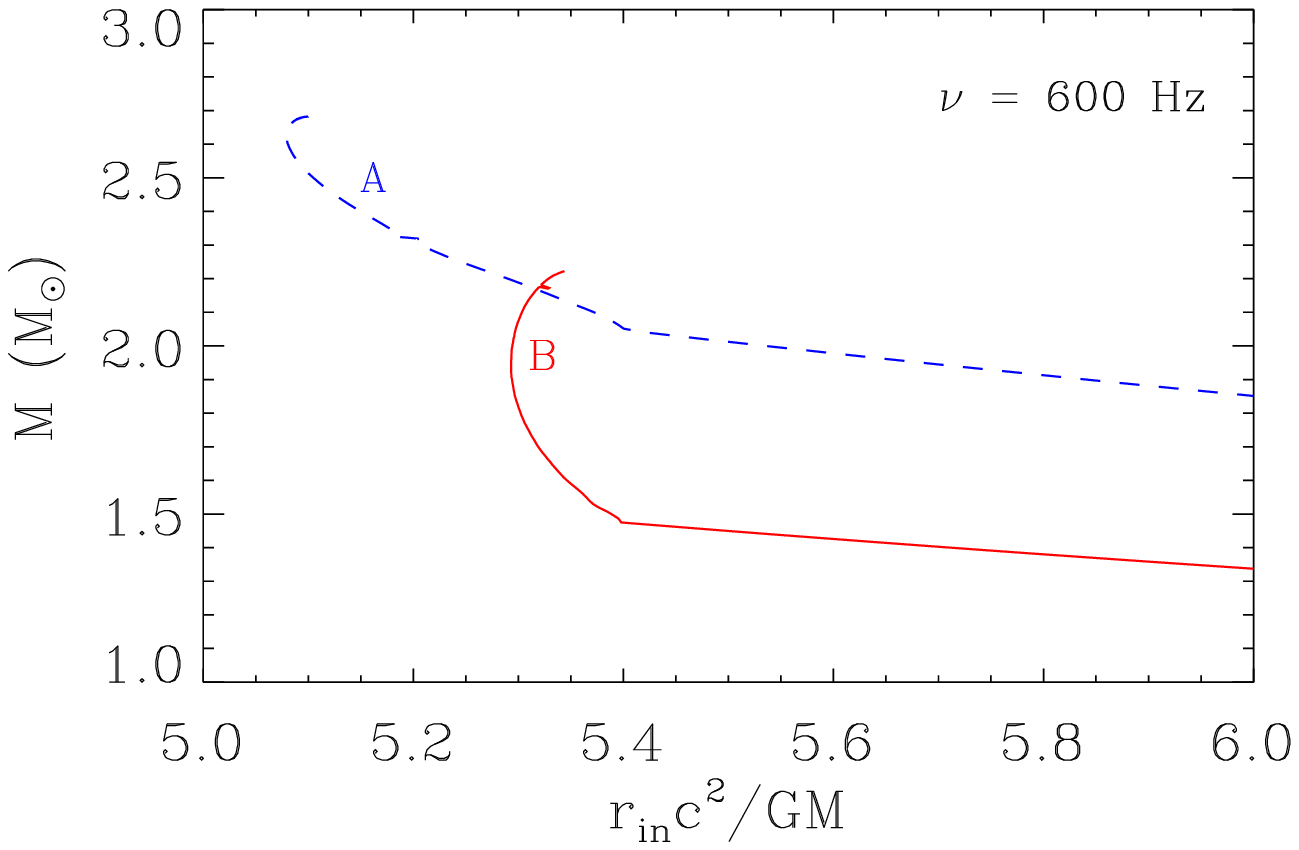}
	\caption{Similar to Figure~\ref{f7}, but for a stellar spin frequency of 600 Hz (Bhattacharyya 2011; see also Section~3).
	\label{f8}}
\end{figure*}

\begin{figure*}
	\centering\includegraphics[height=.35\textheight]{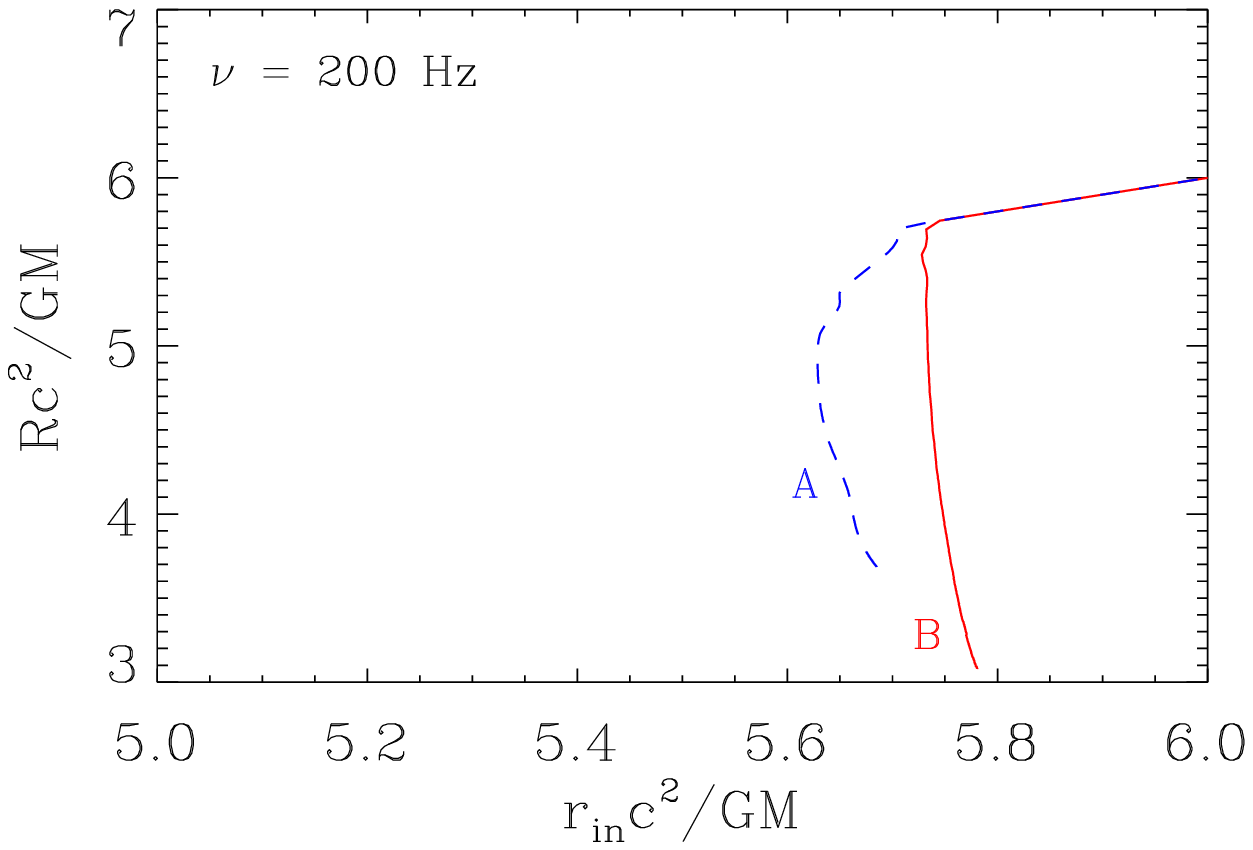}
	\caption{Neutron star radius-to-mass ratio versus disc inner edge radius-to-stellar-mass ratio for
	two realistic EoS models ({\em A} and {\em B}; see Section~3) for a
        stellar spin frequency of 200 Hz (Bhattacharyya 2011).
        The straight line portions of the curves correspond to the disc termination on the stellar surface.
        This figure shows how a measurement of the disc inner edge radius-to-stellar-mass ratio
        from the fitting of a broad relativistic iron K$\alpha$ spectral line can be used to constrain
	neutron star EoS models, when the stellar spin frequency is known.
\label{f9}}
\end{figure*}

\begin{figure*}
	\centering\includegraphics[height=.35\textheight]{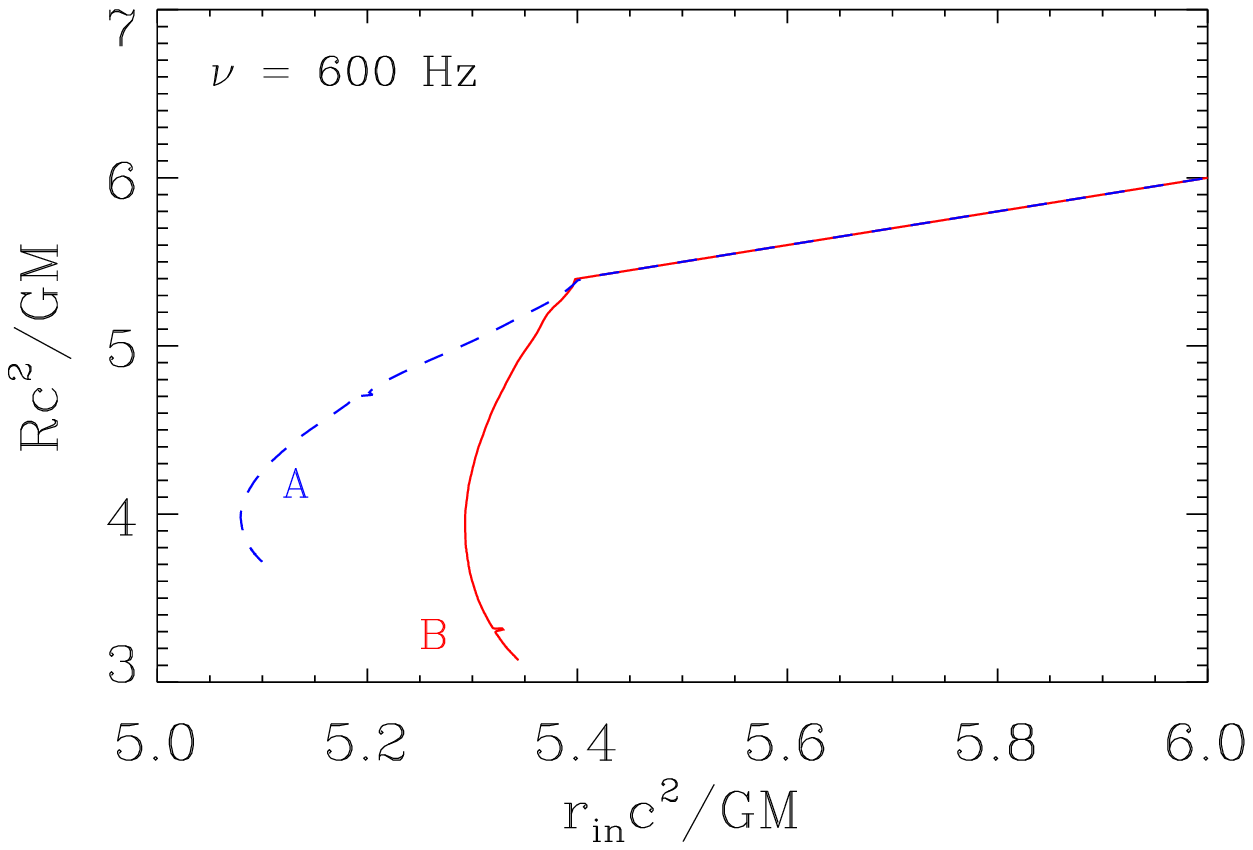}
	\caption{Similar to Figure~\ref{f9}, but for a stellar spin frequency of 600 Hz (Bhattacharyya 2011; see also Section~3).
	\label{f10}}
\end{figure*}

Computation of $r_{\rm in}c^2/GM$ for a neutron star is much more challenging
than that for a black hole, because neutron star has a hard surface. For
the time being, let us assume that the disc is not truncated at a higher radius 
by magnetic field, radiative pressure or other means (Section~1). In such a case, $r_{\rm in}c^2/GM$
can always be obtained from a known analytical expression of $r_{\rm ISCO}$ for a 
black hole (see Section~1). But for a neutron star LMXB, the disc terminates
either at the stellar radius $R$ or at $r_{\rm ISCO}$, whichever is bigger.
Both $R$ and $r_{\rm ISCO}$ depend on the stellar EoS, mass and the spin rate. 
Therefore, while $r_{\rm in}c^2/GM$ 
of a corotating disc always decreases with an increasing $a$ for a black hole, it can either decrease or
increase with an increasing spin rate for a neutron star (see Figure~\ref{f6}). 
In addition, $r_{\rm in}c^2/GM$
depends on stellar mass and EoS. Hence, the computation of $r_{\rm in}c^2/GM$ for a
neutron star requires the calculation of the rapidly spinning stellar structure
considering the full effect of general relativity. This is because the neutron stars
in LMXBs are expected to be rapidly spinning (spin frequency $\nu \sim$~a few hundred Hz),
and hence a non-spinning approximation for stellar structure calculation will not serve 
our purpose, as we are looking for the effects of the stellar spin rate on $r_{\rm in}c^2/GM$.

The method to numerically compute a rapidly spinning neutron star structure, and thus to calculate 
$R$, $r_{\rm in}$ and other stellar parameters for a given EoS model, were detailed
in a number of papers (Bhattacharyya {\em et al.} 2000,
2001a,b,c; Bhattacharyya 2002; Cook {\em et al.} 1994; Datta {\em et al.} 1998),
and we will not repeat it here. In fact, Bhattacharyya (2011) calculated $r_{\rm in}$ for
$\sim 16000$ stable neutron star structures. Here, we review the method mentioned in
Bhattacharyya (2011), as currently this is the only paper that reported a detailed procedure to
constrain EoS models using relativistic disc lines. Here, for the purpose of demonstration,
we use two theoretically proposed nucleonic EoS models, which can support the maximum
observed neutron star mass ($2.01\pm0.04 M_\odot$; Section~1). These EoS models are briefly
described here. (1) {\it Model A}: this is the stiffer EoS model among the two,
and is a field theoretical chiral sigma model for neutron-rich matter in beta equilibrium
(Sahu {\em et al.} 1993). (2) {\it Model B}: this is the Argonne $v_{18}$ model of two-nucleon 
interaction, including the three-nucleon interaction (Urbana IX [UIX] model) and also the 
effect of relativistic boost corrections (Akmal {\em et al.} 1998).

We demonstrate the ways to constrain EoS models for an inferred $r_{\rm in}c^2/GM$ value
using Figures~\ref{f7}--\ref{f10}. We assume a known value of $\nu$ for each case, because
$\nu$ is measured for a number of neutron star LMXBs (Patruno and Watts 2012;
Watts 2012). Note that we consider $r_{\rm in}c^2/GM$ up to $6$ in these figures.
This is because $r_{\rm ISCO}c^2/GM = 6$ for a non-spinning neutron star and $r_{\rm ISCO}c^2/GM < 6$
for $\nu > 0$ for a corotating disc. Therefore, when $r_{\rm in}c^2/GM < 6$, the stellar
spin certainly affects $r_{\rm in}c^2/GM$, which is required to distinguish various
EoS model curves using the inferred $r_{\rm in}c^2/GM$ value (Figures~\ref{f7}--\ref{f10}).
Note that observations indicate $r_{\rm in}c^2/GM < 6$ for many cases, as mentioned in Section~2.

The oblique straight line portions
of the curves in Figures~\ref{f7}--\ref{f10} are for $r_{\rm in} = R$, while the other portions are
for $r_{\rm in} = r_{\rm ISCO}$. These figures show that observed
constraints on $M$ (and/or $Rc^2/GM$) and $r_{\rm in}c^2/GM$ can constrain EoS models.
However, one needs to measure $r_{\rm in}c^2/GM$ typically with better than $0.1$ accuracy in order
to distinguish curves for various EoS models.

While constrining EoS models, if we assume that the inferred $r_{\rm in}c^2/GM$ is exactly 
same as the numerically computed $r_{\rm in}c^2/GM$ (i.e., $Rc^2/GM$ or $r_{\rm ISCO}c^2/GM$,
whichever is greater), then systematics will be introduced in such constraints.
This is because, even when the inferred $r_{\rm in}c^2/GM$ is less than $6$, the disc inner edge radius 
can be affected by stellar magnetic field, radiative pressure or other effects, and the disc
could be truncated at a slightly higher radius value than $R$ and $r_{\rm ISCO}$. 
Therefore, the observationally inferred $r_{\rm in}c^2/GM$ should be taken as the upper limit, while
comparing with the numerical results shown in Figures~\ref{f7}--\ref{f10}. 
Note that EoS models can be effectively constrained even with this upper limit. For example, a suitable value of the
measured upper limit of $r_{\rm in}c^2/GM$ can rule out softer EoS models, even if neither
$M$ nor $Rc^2/GM$ is measured (Figures~\ref{f7}--\ref{f10}). Moreover, if an upper limit of
$M$ is known in addition to the upper limit of $r_{\rm in}c^2/GM$, then some of the stiffer EoS
models could be ruled out (see Figures~\ref{f7}--\ref{f8}).

\section{Conclusion}\label{Conclusion}

Relativistic disc line from neutron star systems is a decade-old field. While this new field is already
well established by many detections of such a line, its utility to constrain neutron star EoS models is
not yet well explored. Here we review the rapid development of this field and its current status. We also
discuss how a relativistic disc line, which provides the accretion disc inner edge radius to stellar mass
ratio, can be used to effectively constrain neutron star EoS models, if this ratio is measured with an 
accuracy better than 0.1. This accuracy could be achieved for a reasonable exposure with a future X-ray satellite,
such as {\em eXTP} and {\em Athena}.

\section*{Acknowledgement}

We thank E. M. Cackett, J. M. Miller, D. Barret and C.-Y. Chiang for providing figures.

\end{document}